\documentclass[english,12pt]{article}
\usepackage[cp1251]{inputenc}
\usepackage{babel}
\usepackage{amsmath, amssymb}
\usepackage{axodraw}
\usepackage{epsfig}
\begin{document}
\title{\bf Production of four charged leptons in electroweak $\gamma\gamma$
interactions} \vspace{0.5cm}
\author{ I. Sotsky \\  \it  NC PHEP ,BSU, Minsk, Belarus}
\date{}
\maketitle
\begin{abstract}

 At present paper we performed a detailed numerical
analysis of four charged leptons  photoproduction in frame of
standard theory of electroweak interaction.  Total and
differential cross sections are obtained by using of the
Monte-Carlo method of numerical integration. Different energies of
initial photons (60-2000 GeV in c.m.s.), definite and averaged
spin states of interacting particles and fixed kinematical cuts
are considered.
\end{abstract}

\section{Introduction}
$ $
 \vspace{-0.5cm}

 \normalsize
In the nearest future are planned to realize high energy
experiments on  linear collider which will have a possibility to
operate in $\gamma\gamma, \gamma e$ mode  \cite{c1}. This new
capability will provide a great advantage in investigation of new
physics phenomenon, new particles research, and detailed study of
non abelian nature of electroweak interaction. However it is
necessary to  take in to account several important features, to
realize such kind experiments successfully.

 First of all, since produced $W^{\pm}$ and Higgs particles ordinarily decay  within  detector  they  could
be observed via it's decay products only,  for example via several
pairs of leptons in final  state.
  Because of high accuracy and relatively clean environment provided by leptonic
  collider,
   the exact computations of all backgrounds, namely:
 $\gamma\gamma\;\rightarrow 2l$,
$\gamma\gamma\;\rightarrow 2l\; + \;photons$ and
$\gamma\gamma\;\rightarrow 4l\;\;$ processes, are required to
study of such type production.

 Secondly, since  initial high energy  $\gamma$ quantums are produced by Compton
backscattering of laser ray on  fast electrons, the exact
information about polarization state and energy  spectrum of
obtained photon's beams is necessary to  interpret  the results of
such experiments correctly.
 This data one  can be gathered from  measuring of several calibration
 processes:
   $\gamma\gamma\;\rightarrow 2l$,
$\gamma\gamma\;\rightarrow 2l\; + \;photons\;\;$ and
$\gamma\gamma\;\rightarrow 4l$ for example.

So it is obviously a great importance in precision investigation
of processes listed behind.  The analysis of
$\gamma\gamma\;\rightarrow 2l$ and $\gamma\gamma\;\rightarrow 2l\;
+ \;photons\;\;$ processes one can find   in \cite{c2}-\cite{c4}
for example. The detailed theoretical description of
$\gamma\gamma\;\rightarrow 4l$ reaction, a lot of useful
references to the works devoted to same problem and expression of
squared matrix element  constructed by helicity amplitude method
one can find in \cite{c5}.
 At this paper we performed the detailed numerical
analysis of two  charged leptons pair production in frame of
standard theory of electroweak interaction. Different energies,
polarization states and definite set of experimental kinematical
cuts are considered.

\vspace{2cm}
\section{Calculation}
$ $
\vspace{-0.5cm}

There are six topologically different Feynman diagrams  describing
process of $\gamma\gamma\rightarrow 4l$ in the lowest order of
standard theory of electroweak interaction (see fig.1).  Whole set
of diagrams can be derived on base of these six ones using C- P-
and crossing symmetries.

$$ $$
 \vspace{2cm}
\begin{center}
\begin{picture}(400,100)
\Photon(10,180)(35,167){1}{2} \Photon(10,100)(35,114){1}{2}
\Line(35,167)(85,175) \Line(85,106)(35,114) \Line(45,150)(35,167)
\Line(35,114)(45,130) \Photon(45,150)(45,130){1}{2}
\Line(85,150)(45,150) \Line(45,130)(85,130)
\Text(15,185)[l]{\small $\gamma(k_1)$} \Text(2,115)[l]{\small
$\gamma(k_2)$}\Text(90,177)[l]{\small
$p_1$}\Text(90,150)[l]{\small $p_2$} \Text(90,130)[l]{\small
$p_3$}\Text(90,104)[l]{\small $p_4$} \Text(35,95)[l]{\small
$(1)$}\Text(7,141)[l]{\small
$W,Z,\gamma$}\Vertex(35,167){2}\Vertex(35,114){2}\Vertex(45,150){2}
\Vertex(45,130){2}

\Photon(150,180)(175,167){1}{2} \Photon(150,100)(175,114){1}{2}
\Line(175,167)(225,175) \Line(225,106)(175,114)
\Line(185,140)(175,167) \Line(175,114)(185,140)
\Photon(185,140)(205,140){1}{2} \Line(225,130)(205,140)
\Line(205,140)(225,150) \Text(155,185)[l]{\small $\gamma(k_1)$}
\Text(150,117)[l]{\small $\gamma(k_2)$}\Text(230,177)[l]{\small
$p_1$}\Text(230,150)[l]{\small $p_3$} \Text(230,130)[l]{\small
$p_4$}\Text(230,104)[l]{\small $p_2$} \Text(175,95)[l]{\small
$(2)$}\Text(185,149)[l]{\small
$Z,\gamma$}\Vertex(175,167){2}\Vertex(175,114){2}\Vertex(205,140){2}
\Vertex(185,140){2}

\Photon(290,180)(315,167){1}{2} \Photon(290,100)(315,114){1}{2}
\Line(315,167)(345,159) \Line(365,106)(315,114)
\Line(315,114)(315,167) \Line(345,159)(365,143)
\Photon(345,159)(355,169){1}{2} \Line(365,158)(355,169)
\Line(355,169)(365,180) \Text(290,185)[l]{\small $\gamma(k_1)$}
\Text(290,117)[l]{\small $\gamma(k_2)$}\Text(370,185)[l]{\small
$p_1$}\Text(370,160)[l]{\small $p_2$} \Text(370,145)[l]{\small
$p_3$}\Text(370,104)[l]{\small $p_4$} \Text(315,95)[l]{\small
$(3)$}\Text(317,172)[l]{\small
$W,Z,\gamma$}\Vertex(315,167){2}\Vertex(315,114){2}\Vertex(345,159){2}
\Vertex(355,169){2}

\end{picture}
\end{center}

\vspace{0cm}

\begin{center}
\begin{picture}(400,100)
\Photon(10,180)(50,140){1}{4} \Photon(10,100)(50,140){1}{4}
\Line(65,155)(85,175) \Line(85,106)(65,125)
\Photon(50,140)(65,155){1}{2}\Photon(50,140)(65,125){1}{2}
\Line(85,150)(65,155) \Line(65,125)(85,130)
\Text(15,185)[l]{\small $\gamma(k_1)$} \Text(-3,117)[l]{\small
$\gamma(k_2)$}\Text(90,177)[l]{\small
$p_1$}\Text(90,150)[l]{\small $p_2$} \Text(90,130)[l]{\small
$p_3$}\Text(90,104)[l]{\small $p_4$} \Text(35,95)[l]{\small
$(4)$}\Text(48,155)[l]{\small $W$}\Text(48,125)[l]{\small
$W$}\Vertex(50,140){2}\Vertex(65,155){2}\Vertex(65,125){2}

\Photon(150,180)(175,160){1}{3} \Photon(150,100)(175,120){1}{3}
\Line(200,160)(225,175) \Line(225,106)(200,120)
\Photon(175,120)(175,160){1}{3}
\Photon(175,120)(200,120){1}{2}\Photon(175,160)(200,160){1}{2}
\Line(200,120)(225,130) \Line(225,150)(200,160)
\Text(155,185)[l]{\small $\gamma(k_1)$} \Text(135,114)[l]{\small
$\gamma(k_2)$}\Text(230,177)[l]{\small
$p_1$}\Text(230,150)[l]{\small $p_2$} \Text(230,130)[l]{\small
$p_3$}\Text(230,104)[l]{\small $p_4$} \Text(175,95)[l]{\small
$(5)$}\Text(183,168)[l]{\small $W$}\Text(183,130)[l]{\small
$W$}\Text(183,168)[l]{\small $W$} \Text(160,142)[l]{\small $W$}
\Vertex(175,160){2}\Vertex(175,120){2}\Vertex(200,120){2}
\Vertex(200,160){2}

\Photon(290,180)(315,167){1}{2} \Photon(290,100)(315,114){1}{2}
\Photon(315,167)(350,167){1}{3} \Line(365,106)(315,114)
\Line(315,114)(335,134) \Line(335,134)(365,134)
\Photon(315,167)(335,134){1}{3} \Line(365,158)(350,167)
\Line(350,167)(365,180) \Text(295,185)[l]{\small $\gamma(k_1)$}
\Text(279,110)[l]{\small $\gamma(k_2)$}\Text(370,185)[l]{\small
$p_1$}\Text(370,160)[l]{\small $p_2$} \Text(370,135)[l]{\small
$p_3$}\Text(370,104)[l]{\small $p_4$} \Text(315,95)[l]{\small
$(6)$}\Text(332,175)[l]{\small $W$}\Text(310,145)[l]{\small
$W$}\Vertex(315,167){2}\Vertex(315,114){2}\Vertex(350,167){2}
\Vertex(335,134){2}

\end{picture}
\end{center}
\vspace{-3.8cm}
\begin{center}
{\small Fig.1. Feynman diagrams for process $\gamma\gamma
\rightarrow 4l$.}
\end{center}

The diagrams  containing charged current exchange are excluded
because only processes with four charged leptons in final state
are considered at present work.

The construction of squared matrix element  is realized both using
helicity amplitude method \cite{c6}-\cite{c9} and precision
covariant one \cite{c10},\cite{c11}. Helicity amplitude method is
used to obtain cross sections  in massless limit at each possible
polarization state. Simple form of final expression allow to
perform numerical integration with high enough accuracy for a
short interval of computing time. The explicit view of  matrix
element constructed by using helicity amplitude method one can
find in \cite{c5}.

The precision covariant method allow to construct squared matrix
element without any approximations at averaged polarization state
of interacting particles. It is applied to verify of received
data, to estimate  error of massless approximation and to
investigate heavy leptons production processes.

The investigation of differential  and  total cross sections  is
realized by using of the Monte-Carlo method of numerical
integration.

\section{Results and Conclusion}
$ $
\vspace{-0.5cm}

 We have calculated the differential and total cross
sections of $\gamma\gamma\;\rightarrow 4l$ process in frame of
standard theory of electroweak interaction at various sets of
energies, polarization states and  kinematical cuts.

The values of total cross sections at different kinematical
conditions  are presented in the Table. The differential cross
section dependences on cosine of polar angle are shown on
figs.2-9. The case of averaged polarization state of interacting
particles as well as fixed ones is considered. The total cross
section dependence on energy of initial  particles one can see on
fig.10.

Presented results demonstrate the cross section's  pronounced
dependence on cosine of polar angle: cross section is almost
equals to zero in the middle part of kinematical region (the
kinematical region where final particles have large $p_{\bot}$)
and extremely increase in the part close to the borders (the
kinematical region with small $p_{\bot}$) (see figs.2-9). Also,
the total cross section strongly increase with decreasing  energy
of interacting photons (see the Table and fig.10). Such behavior
causes because an expression of cross section contain
$\frac{\displaystyle 1}{\displaystyle k_1k_2}$ phactor, where
$k_1$ and $k_2$ are four momenta of initial photons.

The assymetry of differential cross section  (figs.4,5) occurs
because the probability of  lepton  scattering  into direction
close to direction of initial photon is larger in case  of
different signs of photon polarization  and  scattering lepton
helicity then in case of equal ones.

As one can see on figs.6,7 the differential cross section has
symmetric form since  initial  photons have the similar
polarization signs, but the cross section's value  is larger in
case when lepton's helicity sign  is different from polarization
signs of interacting photons. The total cross section of lepton
photoproduction with fixed polarization states  is larger in case
of different polarizations of initial photos then in case of
similar ones (see fig.8,9).

The comparison of results obtained by helicity amplitude method
and precision covariant one indicates that electron and positron
mass contributions have vanishingly small value at investigated
kinematical region. Also, as one can see in the Table, the
distinguish between $\gamma \gamma \rightarrow 4\mu$ cross section
and $\gamma \gamma \rightarrow 4e$ one at the energy of initial
particles $0.5 Tev$ (in c.m.s.) is less then Monte-Carlo
statistical error. So, at the energy of initial photons $0.5 Tev$
(in c.m.s.) and higher one could neglect muon mass contribution
without precision losing. As well, we could perform the same
neglect  in $\gamma \gamma \rightarrow e^+e^-\mu^+\mu^-$ reaction
with interacting energy $0.5 Tev$ (in c.m.s.) and higher by the
similar reason.

The squared matrix element, constructed by  precision covariant
method, is obtained  for unpolarized interacting particles only,
so one could't investigate cross sections at definite spin states
in case of heavy particles production ($\mu$ and $\tau$ leptons).
However, if the interacting energy is larger then value indicate
above (0.5 Tev for $\gamma \gamma \rightarrow
\mu^+\mu^-\mu^+\mu^-$ and  $\gamma \gamma \rightarrow
e^+e^-\mu^+\mu^-$ reactions ), one could  apply the expression of
matrix element constructed by helicity amplitude method for
massless particles and obtain cross section at each possible spin
states.

The relative error of obtained results ($\sim$ 0.5\%-0.9\%) is
less then expected experimental error   in future high energy
experiments on linear collider \cite{c1}.

\begin{center}   Table.

The  total cross sections dependences  on energy of initial
particles. Here the following notation $(\alpha,\beta)$  is used
to describe kinematical cuts, where $\alpha$ is minimal angle
between  any two final particles, $\beta$-- minimal polar angle.
Minimal admissible energy of any final particle is $1 Gev$.
\end{center}

\begin{center}
\begin{tabular}{|c|l|l|l|}
 \hline
 \multicolumn{1}{|c|}{cut} & \multicolumn{3}{|c|}{$(3^o,7^o)$} \\ \hline
\multicolumn{1}{|c|}{energy (Gev)}
&\multicolumn{1}{|c|}{$e^+e^-e^+e^-$}&\multicolumn{1}{|c|}{$\mu^+\mu^-\mu^+\mu^-$}&\multicolumn{1}{|c|}{$e^+e^-\mu^+\mu^-$}
\\ \hline
 $60$&$1254.36\pm 6.73$&$1210.99\pm 8.90$&$630.20\pm 5.49$ \\ \hline
 $120$&$379.86\pm 1.87$&$370.37\pm 2.48$&$189.41\pm 1.55$  \\ \hline
 $200$&$154.94\pm 0.81$&$152.87\pm 1.25$&$78.12\pm 0.54$  \\ \hline
 $300$&$76.41\pm 0.47$&$75.36\pm 0.44$&$38.77\pm 0.23$  \\ \hline
 $400$&$46.97\pm 0.30$&$46.67\pm 0.35$&$23.95\pm 0.17$  \\ \hline
 $500$&$32.15\pm 0.15$&$32.14\pm 0.27$&$16.43\pm 0.14$  \\ \hline
 $1000$&$9.90\pm 0.07$&$9.97\pm 0.12$ &$5.03\pm 0.04$  \\ \hline
 $1500$&$5.05\pm 0.04$&$5.07\pm 0.11$  &$2.55\pm 0.01$   \\ \hline
 $2000$&$3.04\pm 0.03$&$3.03\pm 0.03$  & $1.52\pm 0.01$  \\ \hline

\end{tabular}
\end{center}

\vspace{1cm}

\begin{figure}[ht!]
\leavevmode
\begin{minipage}[b]{.5\linewidth}
\centering
\includegraphics[width=\linewidth, height=7.5cm, angle=0]{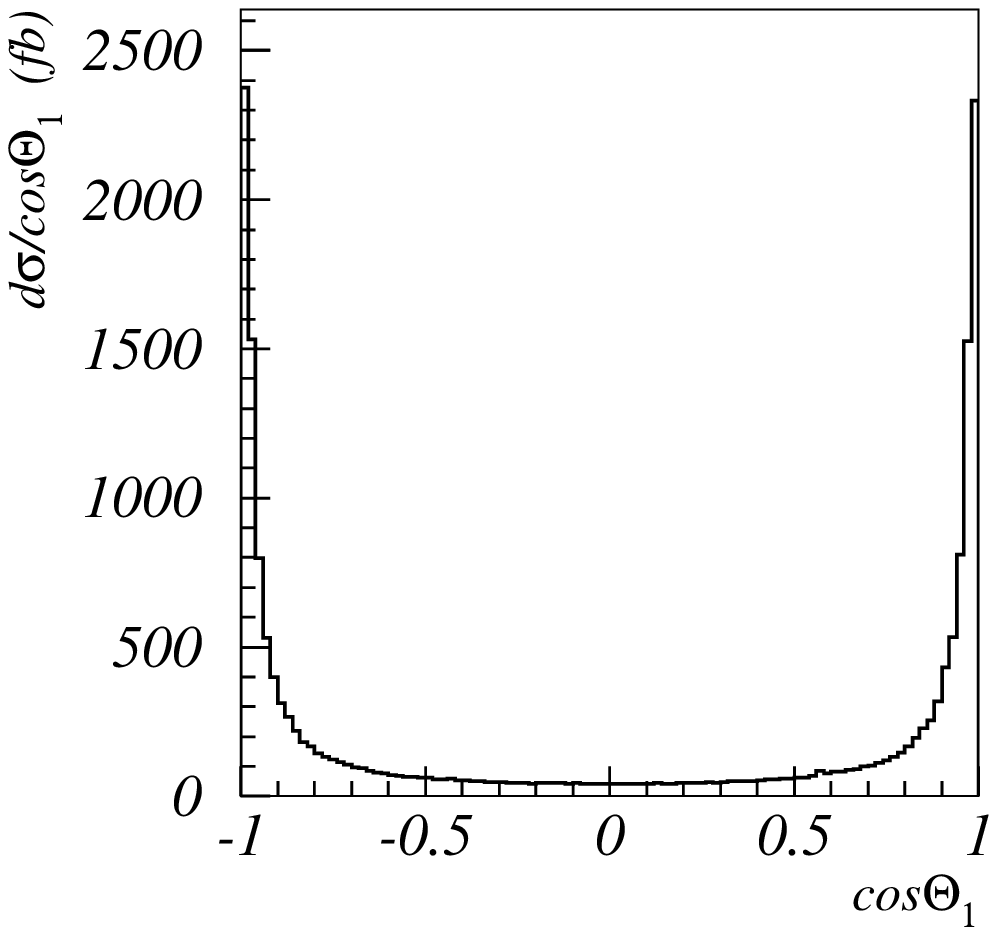}
\end{minipage}\hfill
\begin{minipage}[b]{.5\linewidth}
\centering
\includegraphics[width=\linewidth, height=7.5cm, angle=0]{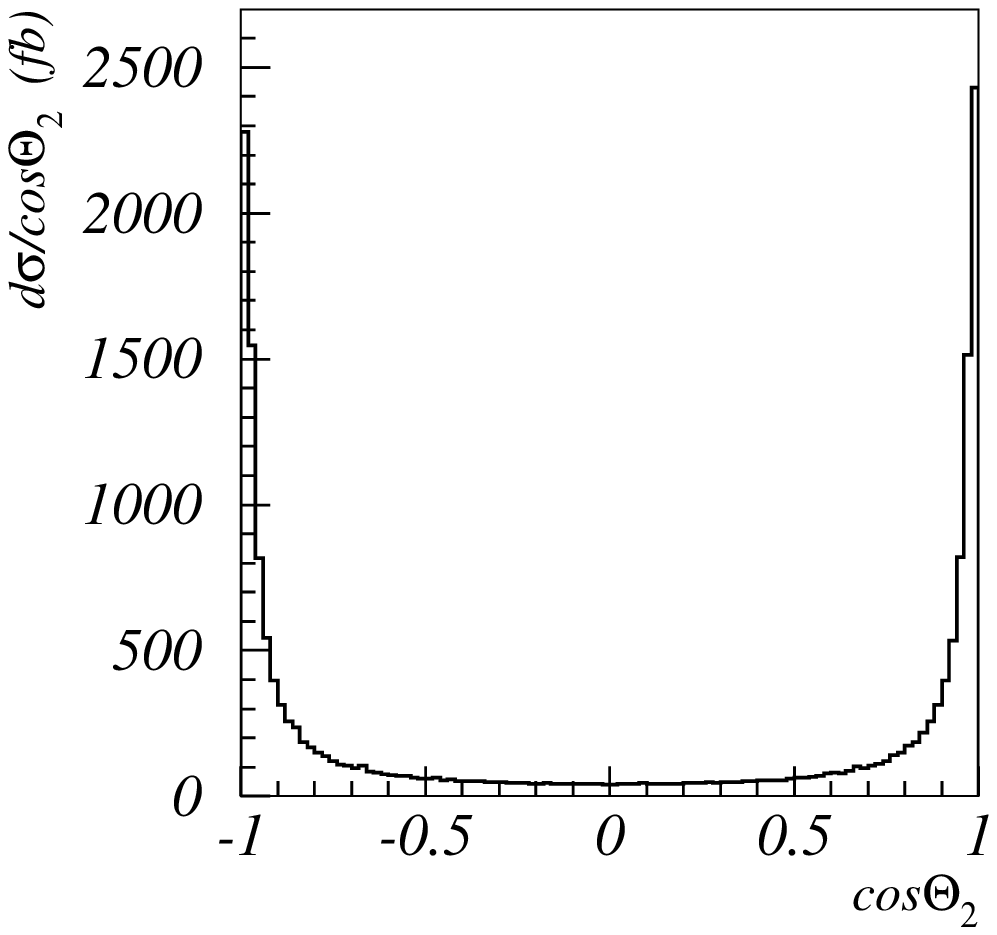}
\end{minipage}
 \vspace{-15pt}
\begin{center}
 { Fig.2. The differential
cross section dependence of $\gamma\gamma \rightarrow
e^+e^-e^+e^-$ process on cosine of polar angle  at averaged
polarization state of interacting particles. Here  energy of
$\gamma\gamma-$ beam  is $ 120 GeV$ in c.m.s, $\theta_{1(2)}$ is
the angle between  the first(second) photon and one of the final
electron. The values of polar angle cut and cut on angle between
any two final particles are $7^o$ and $3^o$ respectively. Minimal
admissible energy of any final particle is $1 Gev$}
 \end{center}
\end{figure}

\vspace{-1cm}

\begin{figure}[ht!]
\leavevmode
\begin{minipage}[b]{.5\linewidth}
\centering
\includegraphics[width=\linewidth, height=7.5cm, angle=0]{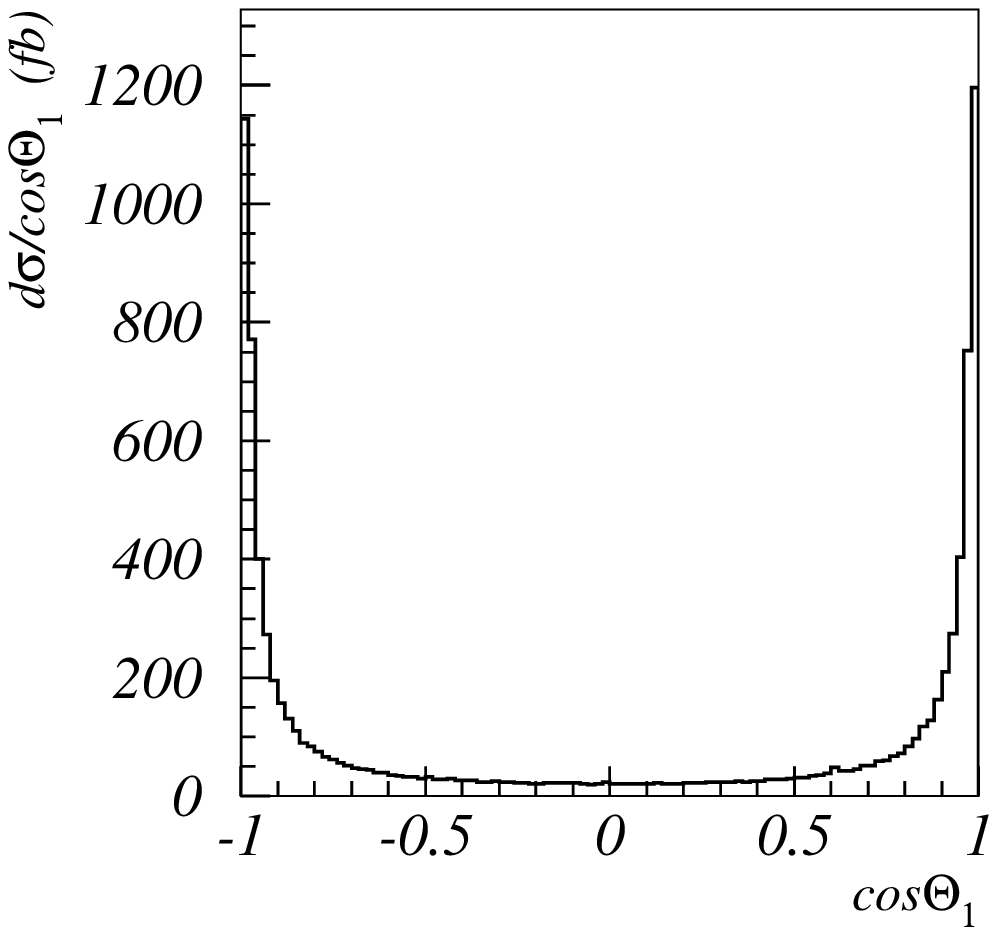}
\end{minipage}\hfill
\begin{minipage}[b]{.5\linewidth}
\centering
\includegraphics[width=\linewidth, height=7.5cm, angle=0]{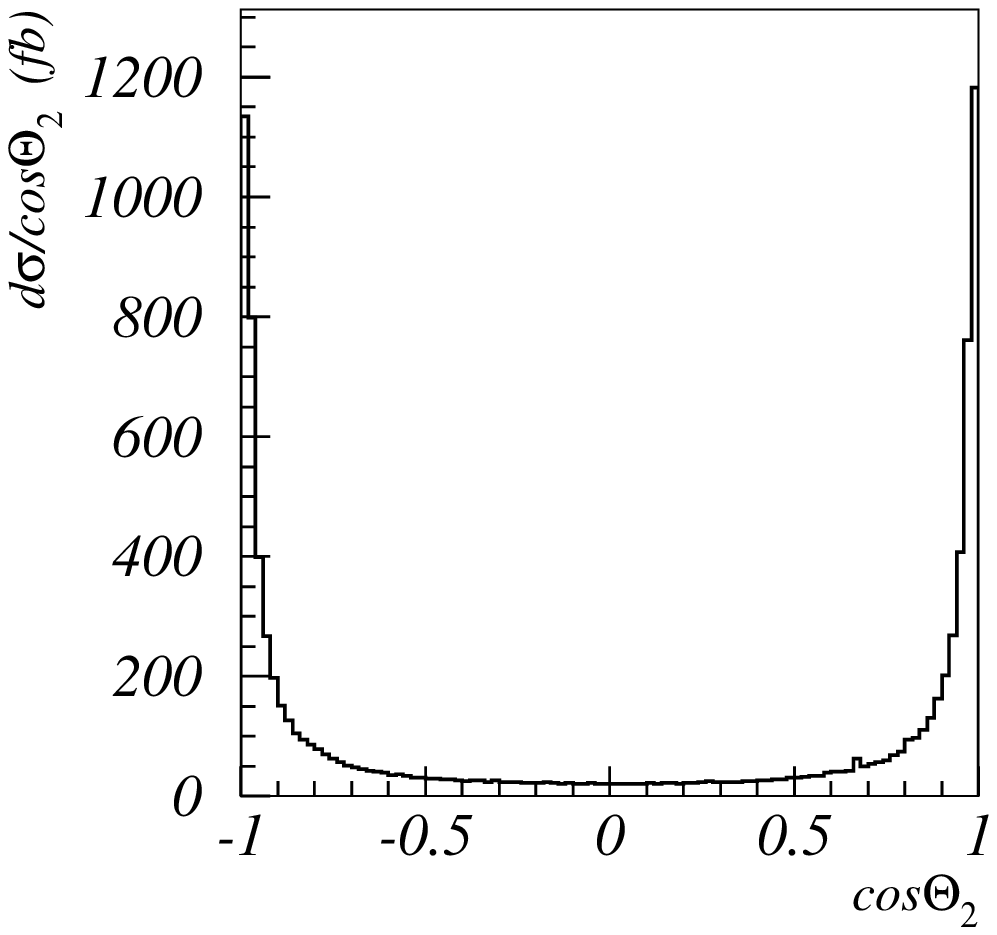}
\end{minipage}
 \vspace{-15pt}
\begin{center}
{ Fig.3. The differential cross section dependence of
$\gamma\gamma \rightarrow e^+e^-\mu^+\mu^-$ process on cosine of
polar angle  at averaged polarization state of interacting
particles. Here   energy of $\gamma\gamma-$ beam  is $ 120 GeV$ in
c.m.s, $\theta_{1(2)}$ is the angle between  the first(second)
photon and one of the final electron. The values of polar angle
cut and cut on angle between any final two particles are $7^o$ and
$3^o$ respectively. Minimal admissible energy of any final
particle is $1 Gev$}
 \end{center}
\end{figure}

\vspace{-0.7cm}
\begin{figure}[ht!]
\leavevmode
\begin{minipage}[b]{.33\linewidth}
\centering
\includegraphics[width=\linewidth, height=7.5cm, angle=0]{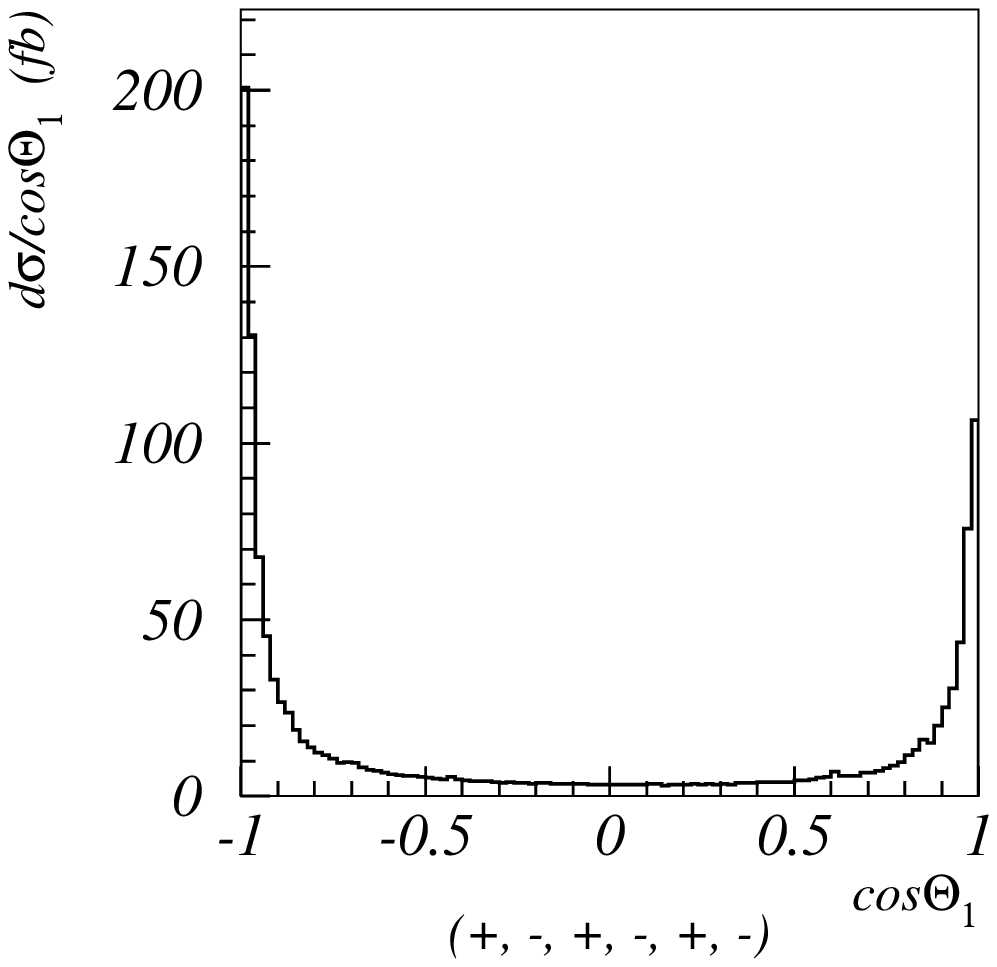}
\end{minipage}
\begin{minipage}[b]{.33\linewidth}
\centering
\includegraphics[width=\linewidth, height=7.5cm, angle=0]{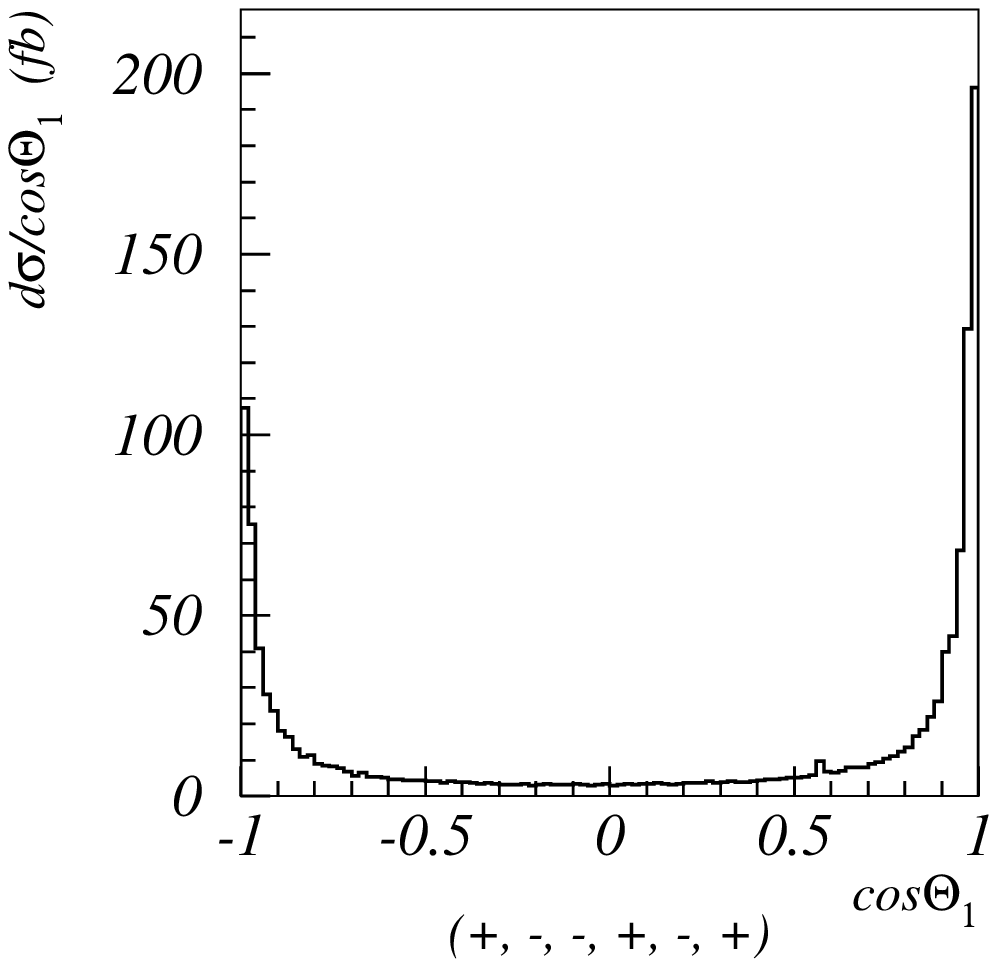}
\end{minipage}\hfill
\begin{minipage}[b]{.33\linewidth}
\centering
\includegraphics[width=\linewidth, height=7.5cm, angle=0]{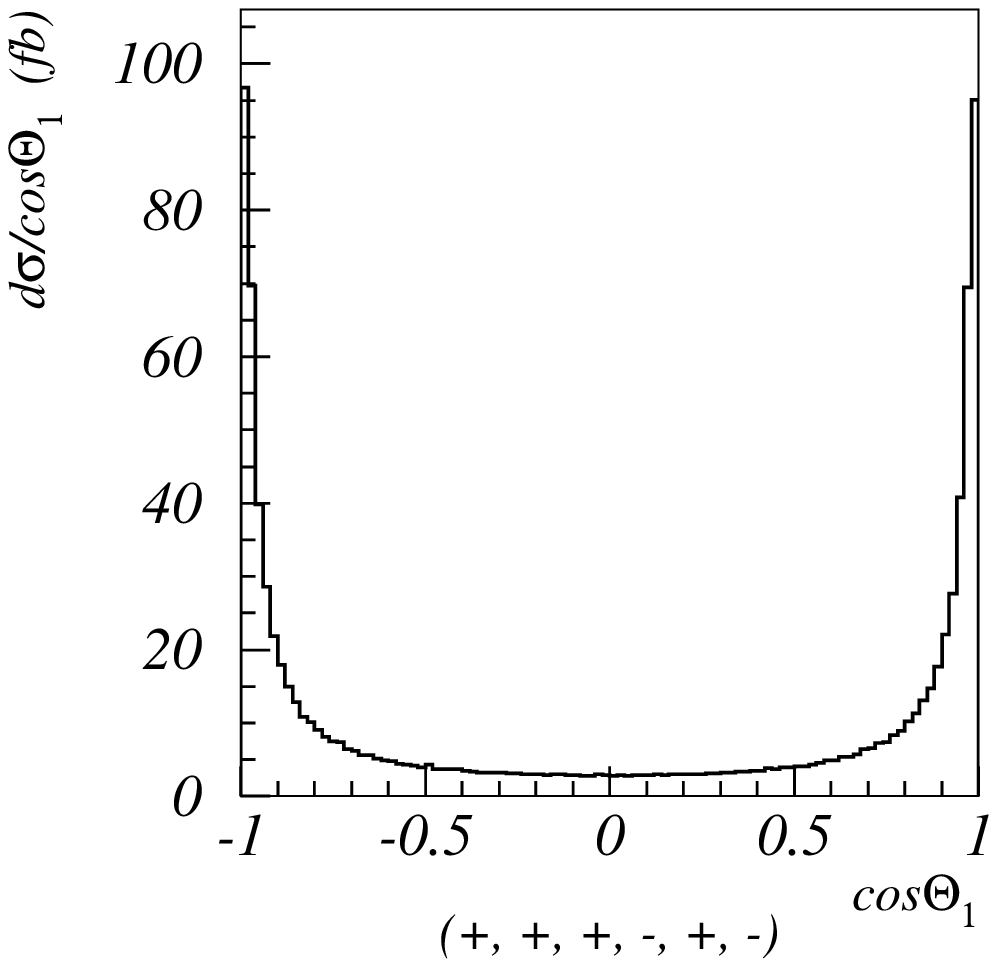}
\end{minipage}

 \vspace{-5pt}

 {\hspace{2.5cm}\small Fig.4. \hspace{4.7 cm} Fig.6. \hspace{4.7 cm} Fig.6.}
\end{figure}

\vspace{-0.7cm}

\vspace{-0.7cm}
\begin{figure}[ht!]
\leavevmode
\begin{minipage}[b]{.33\linewidth}
\centering
\includegraphics[width=\linewidth, height=7.5cm, angle=0]{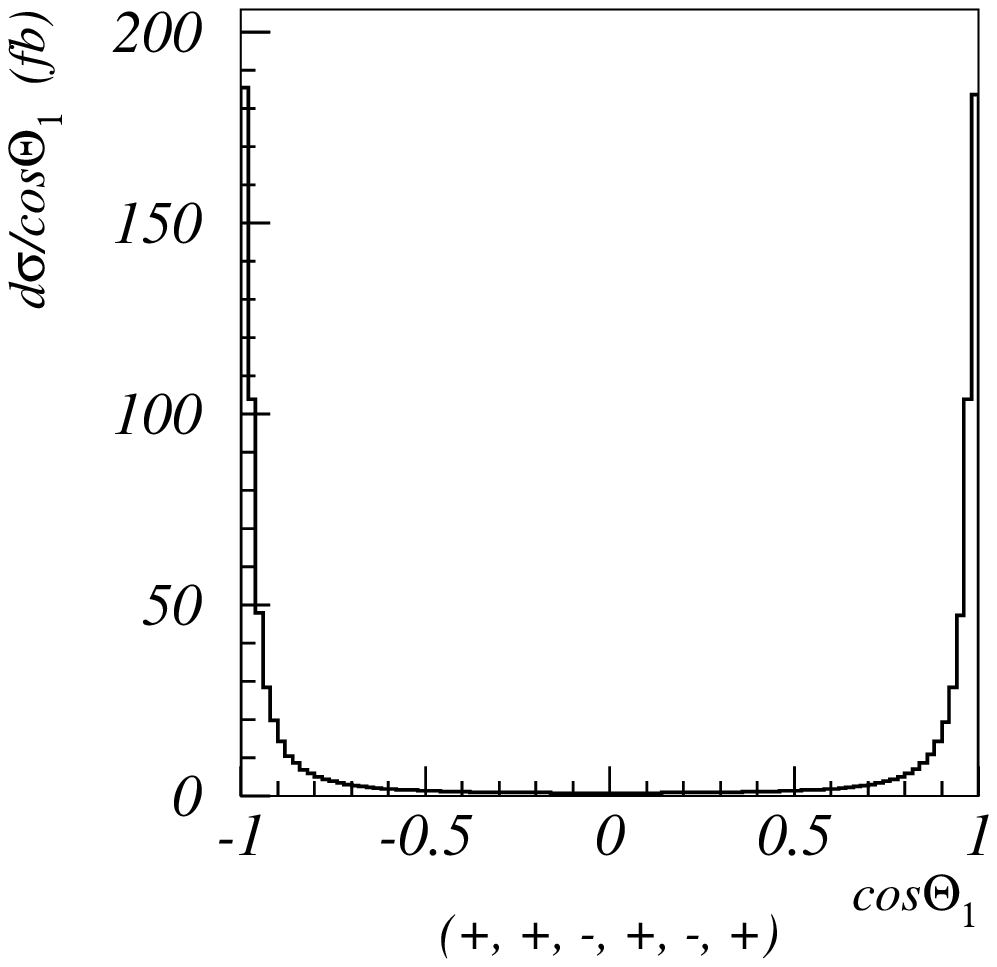}
\end{minipage}
\begin{minipage}[b]{.33\linewidth}
\centering
\includegraphics[width=\linewidth, height=7.5cm, angle=0]{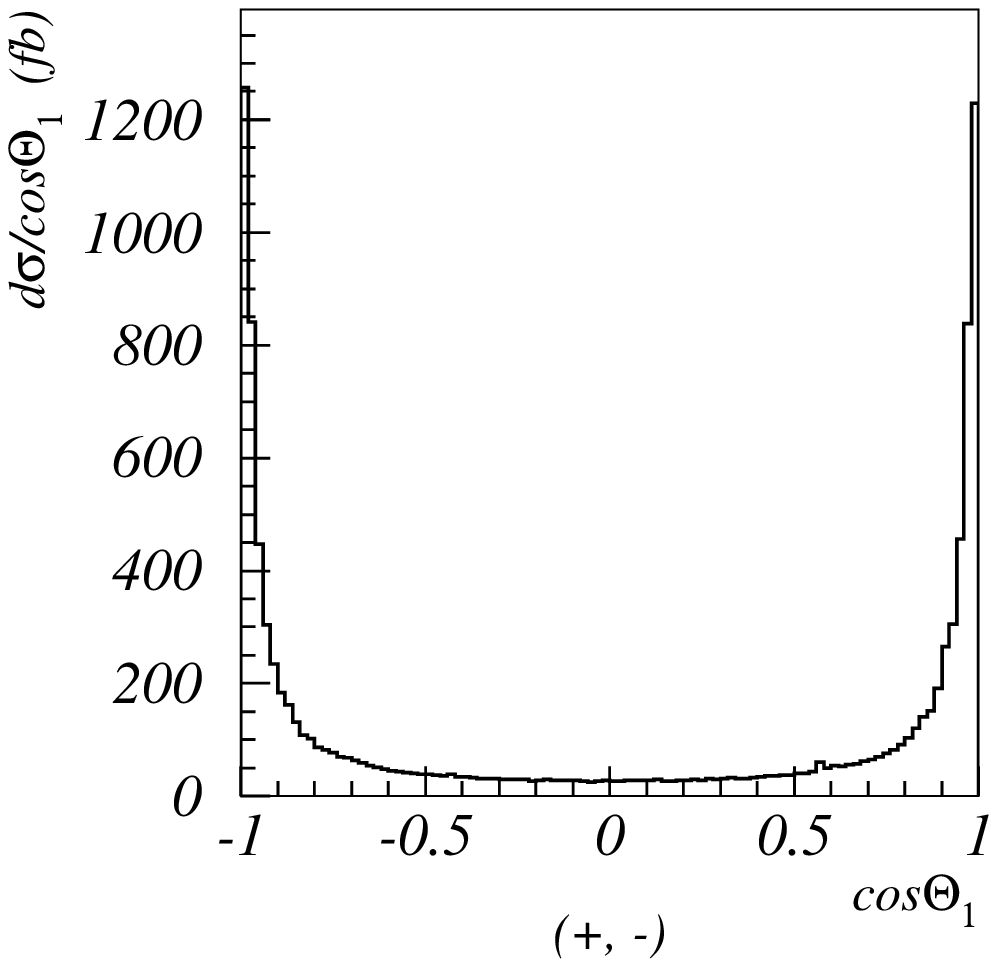}
\end{minipage}\hfill
\begin{minipage}[b]{.33\linewidth}
\centering
\includegraphics[width=\linewidth, height=7.5cm, angle=0]{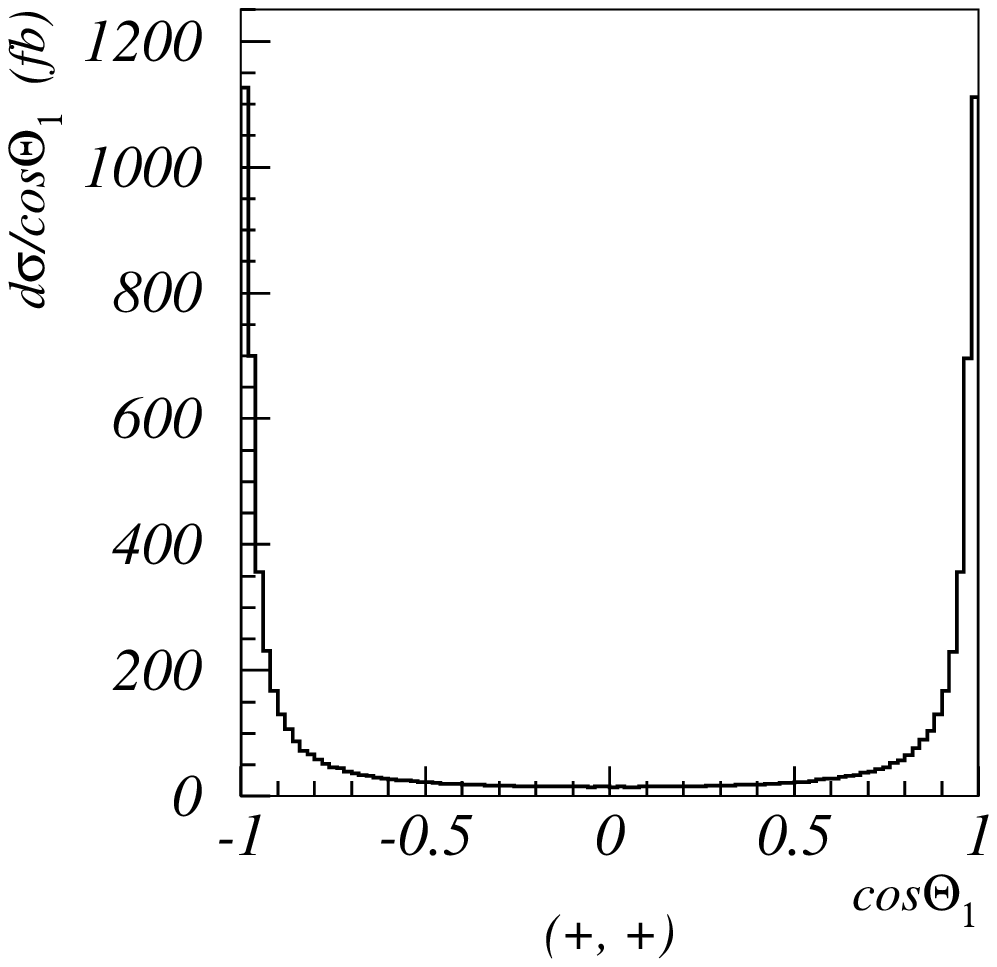}
\end{minipage}
 \vspace{-5pt}
 {\hspace{2.5cm}\small Fig.7. \hspace{4.7 cm} Fig.8. \hspace{4.7 cm} Fig.9.}

\begin{center}
 {
Figs.4-8. The  dependence  of the $\gamma\gamma \rightarrow
e^+e^-e^+e^-$ differential cross section on cosine of polar angle
at fixed polarization states of interacting particles.
 Here  energy of $\gamma\gamma-$ beam  is $ 120 GeV$ in c.m.s, $\theta_{1}$ is the angle between  the
first photon and one of the final electron. The values of polar
angle cut and cut on angle between any two final particles are
$7^o$ and $3^o$ respectively. Minimal admissible energy of any
final particle is $1 Gev$
 For identification of particle polarizations is used following
notation: $(+,-,+,-,+,-,) =
(\lambda_1,\lambda_2,\lambda_3,\lambda_4,\lambda_5,\lambda_6)$},
where $\lambda_{1(2)}\;$ corresponds to polarization state of
photon with four momentum $k_{1(2)}$, $\lambda_{3,4,5,6}\;-$
helicity of lepton with four momentum $p_{1,2,3,4}$\;. Notations
of $(+,-)$ and $(+,-)$  mean spin states with definite photon's
and averaged lepton's polarizations.
\end{center}

\end{figure}

\begin{figure}[ht!]
\leavevmode
\begin{minipage}[b]{1.\linewidth}
\centering
\includegraphics[width=\linewidth, height=10cm, angle=0]{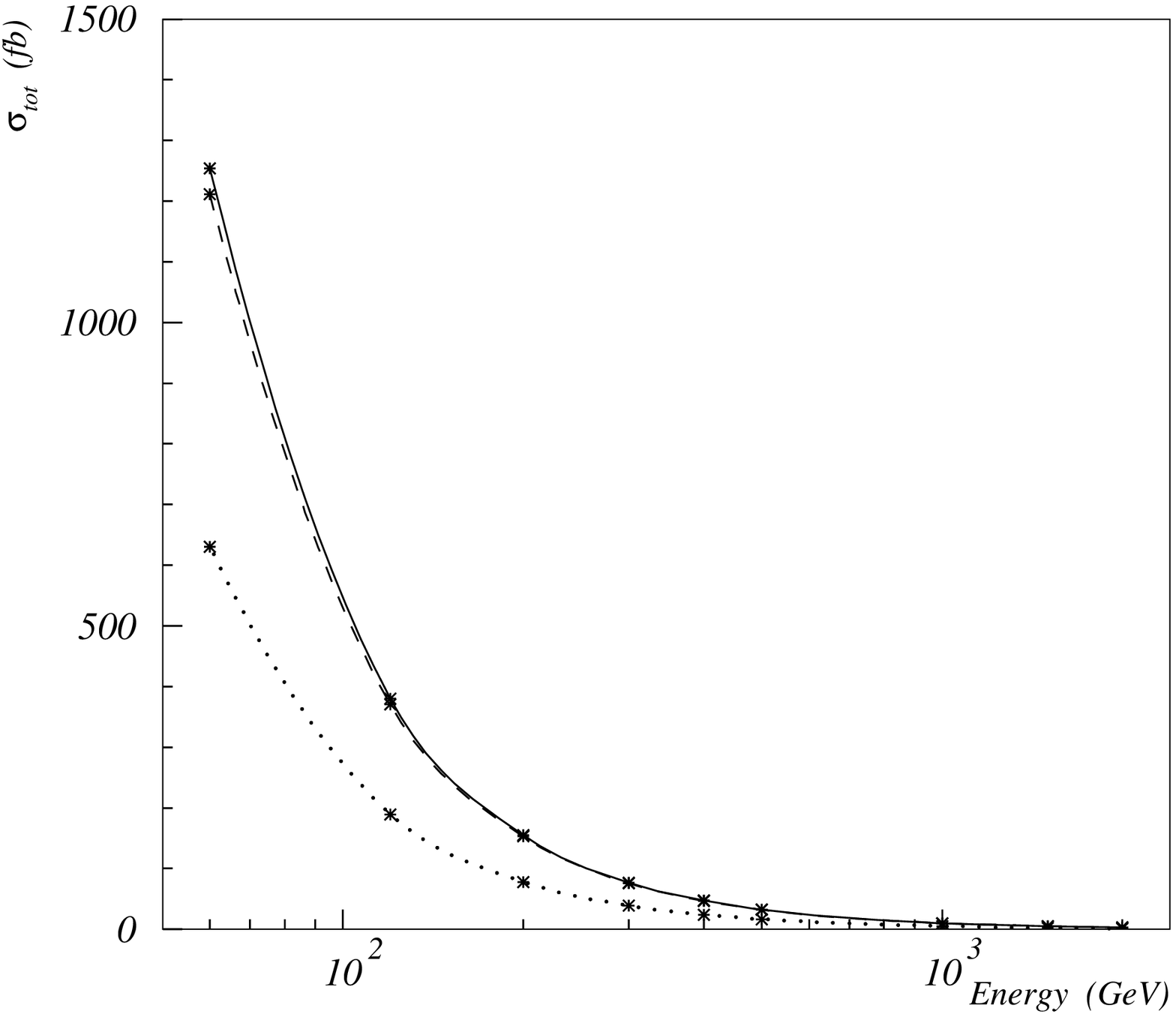}
\end{minipage}
 \vspace{-15pt}
\begin{center}
 {Fig.10. The total cross sections dependences on energy of interacting
 particles.
Solid line corresponds to cross section of $\gamma\gamma
\rightarrow e^+e^-e^+e^-$ process, dashed line  to $\gamma\gamma
\rightarrow \mu^+\mu^-\mu^+\mu^-$ process and dotted line to
$\gamma\gamma \rightarrow e^+e^-\mu^+\mu^-$ process.}
\end{center}
\end{figure}

\newpage $$ $$\newpage $$ $$ 
$$ $$ $$ $$

\end{document}